# Numerical Simulation of Oscillating Multiphase Heat Transfer in Parallel plates using Pseudopotential Multiple-Relaxation-Time Lattice Boltzmann Method

**Wandong Zhao**
Sch. of Mechanical and
Electrical Engineering,
Nanchang University,
Nanchang, Jiangxi, China

**Ben Xu\***
Depart. of Mechanical
Engineering, The University of
Texas Rio Grande Valley,
Edinburg, TX, USA
ben.xu@utrgv.edu

**Ying Zhang**
Sch. of Mechanical and
Electrical Engineering,
Nanchang University,
Nanchang, Jiangxi, China

**ABSTRACT**

Multiphase flows frequently occur in many important engineering and scientific applications, but modeling of such flows is a rather challenging task due to complex interfacial dynamics between different phases, let alone if the flow is oscillating in the porous media. Using humid air as the working fluid in the thermoacoustic refrigerator is one of the research focus to improve the thermoacoustic performance, but the corresponding effect is the condensation of humid air in the thermal stack. Due to the small sized spacing of thermal stack and the need to explore the detailed condensation process in oscillating flow, a mesoscale numerical approach need to be developed. Over the decades, several types of Lattice Boltzmann (LB) models for multiphase flows have been developed under different physical pictures, for example the color-gradient model, the Shan-Chen model, the nonideal pressure tensor model and the HSD model. In the current study, a pseudopotential Multiple-Relaxation-Time (MRT) LBM simulation was utilized to simulate the incompressible oscillating flow and condensation in parallel plates. In the initial stage of condensation, the oscillating flow benefits to accumulate the saturated vapor at the exit regions, and the velocity vector of saturated vapor clearly showed the flow over the droplets. It was also concluded that if the condensate can be removed out from the parallel plates, the oscillating flow and condensation will continuously feed the cold surface to form more water droplets. The effect of wettability to the condensation was discussed, and it turned out that by increasing the wettability, the saturated water vapor was easier to condense on the cold walls, and the distance between each pair of droplets was also strongly affected by the wettability. It's expected that this study can be used to optimize and redesign the structure of thermal stack in order to produce more condensed water, also this multiphase approach can be extended to more complicated 3D structures.

1. **Introduction**

Understanding the multiphase heat transfer in porous media is of great significance to heat transfer community, because it is involved in many engineering applications, such as oil reservoir engineering, enhanced geothermal energy and ground water contamination or remediation. Another potential application is the thermoacoustic refrigerator using moisturized air as working fluid. Hiller & Swift [1] discussed the condensation in an open-flow thermoacoustic cooler with stack temperatures below the saturation temperature of the humid air, and they concluded that the condensation didn't significantly affect the oscillating thermoacoustic, but the condensing water was a large thermal load on the refrigerator. Slaton *et. al.* [2] also explored the "wet" thermoacoustic effect using condensable gas as the working fluid, they concluded that the "wet-walled" system can improve the performance of thermoacoustic refrigerators by condensing the gas in the thermal stack.

Consequently, the optimal structure of thermal stack or "regenerator" plays an important role in the operation of thermoacoustic cooler to reach the maximum condensation rate. Nevertheless, the analysis of oscillating flow and condensation in Ref. [1] and [2] are mainly replying on the linear theory of thermoacoustic, proposed by Rott [3] and Swift [4], a detailed analysis about the condensation of water droplets in the oscillating flow has not been performed yet.

Because of the small spacing of thermal stack in thermoacoustic cooler, it represents a great challenge to explore the multiphase heat transfer from macroscale perspective, for example Yu *et. al.* [5] clearly demonstrated in their study that using the porous media model in ANSYS Fluent led to a failure numerical simulation of thermoacoustic standing-wave engine. Consequently, a mesoscale numerical approach can be utilized to explore the condensation process in thermoacoustic devices.

Over the past decades, several types of multiphase Lattice Boltzmann (LB) models have been developed, for example the first multiphase LB model is the color-gradient model proposed by Gunstensen et al. [6], then Shan and Chen [7, 8] proposed another type of multiphase LB model, known as the pseudopotential model, by introducing an artificial interparticle potential to describe fluid interactions. With the idea of free energy, Swift et al. [9] constructed a third type of multiphase LB model in which a nonideal pressure tensor related to the free-energy functional was introduced. He-Shan-Doole (HSD) model [10, 11] provided a solid theoretical foundation for a transition from the continuous Boltzmann equation to the lattice Boltzmann equation. Nevertheless, all these multiphase models adopted the Bhatnagar-Gross-Krook (BGK) approximation and single-relaxation time (SRT) model [12], therefore those models have some disadvantages in numerical stability because of the spurious flow around the interface of multiphase flow and the inability to simulate problems with Pr number other than order of unity [13]. Furthermore, to the authors' best knowledge, the multiphase LBM models have not been applied to oscillating flow and heat transfer yet.

In the current study, a pseudopotential Multiple-Relaxation-Time (MRT) LBM simulation was adopted to simulate the incompressible oscillating flow and condensation in porous media with simple 2D geometry, specifically inside a 2D channel with flat plates on top and bottom. To improve the numerical stability, the MRT collision model was used instead of the single-relaxation-time (SRT) collision model, by separating the relaxation rates of the hydrodynamic and non-hydrodynamic moments. The model was first verified for a single droplet evaporation and compared with the $D^2$ law, and then a continuous flow was adopted to further verify the model, by simulating the condensation on cold surface with gravity. Finally, an oscillating multiphase flow and heat transfer in two parallel plates was simulated, where the saturated water vapor was condensing on the hydrophobic surface of plates. It turned out that if the condensate can be removed out from the channel, the oscillating flow and heat transfer will not be influenced. The effect of wettability to the condensation was also discussed. It's expected that this study can be used to optimize and redesign the structure of thermal stack in order to produce more condensed water, also this multiphase approach can be extended to more complicated 3D structures.

## 2. Numerical model

### 2.1 Pseudopotential MRT-LB model for fluid flow

The pseudopotential model proposed by Shan-Chan [7, 8] can be used to achieved the multiphase fluid interaction. However, there are some drawbacks because of the Bhatanagar-Gross-Krook (BGK) collision operator [14]. In recent years, Multi-Relaxation-Time (MRT) collision model was used in the Lattice Boltzmann method because of its better numerical stability [15]. With the MRT model, the evolution of the density distribute function (DF) can be obtained as [16, 17]:

$$f_i(x+e_i\delta_t, t+\delta_t) - f_i(x,t) = -M^{-1}S[m(x,t) - m^{eq}(x,t)] \tag{1}$$

where $f$ and $f^{eq}$ represent the particle DF and the equilibrium DF respectively, $\delta x$ and $\delta t$ are the lattice space step and the time step, and both were set to be equal to 1, so $c = \delta x / \delta t = 1$ [18]. $e_a$ denotes the discrete velocity along the direction of $a$. In this study, the D2Q9 model was adopted, so the discrete velocity can be given as [19]:

$$e_i = \begin{cases} (0,0), & i = 0 \\ (1,0)c,(0,1)c,(-1,0)c,(0,-1)c & i = 1-4 \\ (1,1)c,(-1,1)c,(-1,-1)c,(1,-1)c, & i = 5-8 \end{cases} \tag{2}$$

As the same time, $\bar{\Lambda} = M^{-1} \Lambda M$ is the collision matrix, $M$ is the orthogonal transfer matrix, and $\Lambda$ is the diagonal relaxation matrix in the moment space, which can be defined as [20, 21]:

$$\begin{aligned} \Lambda &= diag(s_0, s_1, s_2, s_3, s_4, s_5, s_6, s_7, s_8) \\ &= diag(\tau_\rho^{-1}, \tau_e^{-1}, \tau_\varsigma^{-1}, \tau_j^{-1}, \tau_q^{-1}, \tau_j^{-1}, \tau_q^{-1}, \tau_\upsilon^{-1}, \tau_\upsilon^{-1}) \end{aligned} \tag{3}$$

where $s_1 = s_2$, $s_3 = s_5$, $s_7 = s_8$. The flow non-dimensional relaxation time is defined:

$$\tau_\upsilon = \frac{1}{s_7} = \upsilon / c_s^2 + 0.5 \tag{4}$$

where $\upsilon$ is the kinematic viscosity of the fluid.

The viscosity relaxation time in the computational is determined by:

$$\tau_\upsilon = \tau_g + \frac{\rho - \rho_g}{\rho - \rho_l}(\tau_l - \tau_g) \tag{5}$$

where subscripts $g$ and $l$ denote the gas phase and liquid phase, respectively. By applying linear transformation, $f$ can be converted to the moment space $m = M \cdot f$, $m^{eq} = m \cdot f^{eq}$. The moment space DF $m$ and the equilibrium DF $m^{eq}$ are determined by [15]:

$$\begin{bmatrix} m_0(\rho) \\ m_1(e) \\ m_2(\varepsilon) \\ m_3(j_x) \\ m_4(q_x) \\ m_5(j_y) \\ m_6(q_y) \\ m_7(p_{xx}) \\ m_8(p_{xy}) \end{bmatrix} = \begin{bmatrix} 1 & 1 & 1 & 1 & 1 & 1 & 1 & 1 & 1 \\ -4 & -1 & -1 & -1 & -1 & 2 & 2 & 2 & 2 \\ 4 & -2 & -2 & -2 & -2 & 1 & 1 & 1 & 1 \\ 0 & 1 & 0 & -1 & 0 & 1 & -1 & -1 & 1 \\ 0 & -2 & 0 & 2 & 0 & 1 & -1 & -1 & 1 \\ 0 & 0 & 1 & 0 & -1 & 1 & 1 & -1 & -1 \\ 0 & 0 & -2 & 0 & 2 & 1 & 1 & -1 & -1 \\ 0 & 1 & -1 & 1 & -1 & 0 & 0 & 0 & 0 \\ 0 & 0 & 0 & 0 & 0 & 1 & -1 & 1 & -1 \end{bmatrix} \begin{bmatrix} f_0 \\ f_1 \\ f_2 \\ f_3 \\ f_4 \\ f_5 \\ f_6 \\ f_7 \\ f_8 \end{bmatrix}, m^{(eq)} = \begin{bmatrix} \rho \\ -2\rho + 3(j_x^2 + j_y^2) \\ \rho - 3(j_x^2 + j_y^2) \\ j_x \\ -j_x \\ j_y \\ -j_y \\ j_x^2 - j_y^2 \\ j_x j_y \end{bmatrix} \tag{6}$$

where $j_x$ and $j_y$ are equal to $\rho u_x$, $\rho u_y$, respectively.

In the MRT-LB model, the collision is calculated in the moment space, while the streaming process is carried out in the velocity space. The right-hand side of Eq. (1) can be converted as follows [22]:

$$m^* = m - \Lambda(m - m^{eq}) + \delta_t (I - \frac{\Lambda}{2})\bar{S} \tag{7}$$

where $I$ is the unit tensor, $\bar{S} = MS$ is the forcing term in the moment space, and $S = (S_0, S_1, S_2, S_3, S_4, S_5, S_6, S_7, S_8)^T$.

In order to realize the thermodynamic consistency, Li [20] fourthly improved the source term in Eq. (7), which can be defined in the moment space as:

$$\bar{S} = \begin{bmatrix} 0 \\ 6(u_x F_x + u_y F_y) + \dfrac{12\theta |F_m|^2}{\psi^2 \delta_t (\tau_e - 0.5)} \\ -6(u_x F_x + u_y F_y) + \dfrac{12\theta |F_m|^2}{\psi^2 \delta_t (\tau_\varsigma - 0.5)} \\ F_x \\ -F_x \\ F_y \\ -F_y \\ 2(u_x F_x - u_y F_y) \\ (u_x F_y - u_y F_x) \end{bmatrix} \tag{8}$$

where $\theta$ is used to maintain the numerical stability, and $F_m = (F_{mx}, F_{my})$ is the force acting on the pseudopotential model.

In the pseudopotential model, Shan-Chen [7] employed the interaction particle force, which is the key to mimic the multiphase separation. For the fluid-fluid interaction force in the Shan-Chen model for multiphase flow can be given by:

$$F_m = -G\psi(x,t) \left[ \sum_i w(|e_i|^2) \psi(x + e_i, t) e_i \right] \tag{9}$$

where $G$ is the interaction strength with a positive (negative) sign for a repulsive (attractive) force between particles, and $w(|e_a|^2)$ is the weight factor [21, 22]. For the case of D2Q9 lattice, the weights factors are $w(1) = 1/3$ and $w(2) = 1/12$. and $\psi$ in Eq. (11) can be defined as [23]:

$$\psi = \sqrt{\frac{2(P_{EOS} - \rho c_s^2)}{Gc^2}} \tag{10}$$

where $P_{EOS}$ is non-ideal equation of state. In this paper, the Peng-Robinson equation of state was employed, which can be written as [23]:

$$P_{EOS} = \frac{\rho RT}{1 - b\rho} - \frac{a\varphi(T)\rho^2}{1 + 2b\rho - b^2\rho^2} \tag{11}$$

where $\varphi(T) = [1 + (0.37464 + 1.54226\omega - 0.26992\omega^2)(1 - \sqrt{T/T_c})]^2$, $a = 0.45724 R^2 T_c^2 / P_c$, and $b = 0.0778 RT_c / P_c$. $T_c$ and $P_c$ are the critical temperature and critical pressure, respectively. In this paper, $a$, $b$ and $R$ in Eq.(11) were set to be $a=3/49$, $b=2/21$, $R=1.0$, respectively. Therefore, the critical temperature can be calculated as $T_c=0.1096$.

The streaming process can be expressed as:

$$f_i(x+e_i\delta_t,t+\delta_t) = f_i^*(x,t) \tag{12}$$

where $f^* = M^{-1}m^*$. For the MRT-LB model, the corresponding macroscopic density and velocity are determined by:

$$\rho = \sum_i f_i, \quad \rho u = \sum_i e_i f_i + \frac{\delta_t F}{2} \tag{13}$$

where $F=(F_x,F_y)$ is the total force, which also contain the adhere wall force $F_{ads}$. The contact angle (CA) can be realized by the follows equation [24]:

$$F_{ads} = -G_w \psi(x,t)\left[\sum_i w(|e_i|^2)\psi(\rho_w)s(x+e_i)e_i\right] \tag{14}$$

where $s(x+e_i)$ is a switch function, which is equal to 0 and 1 for solid and fluid, respectively.

The gravitational force $F_g$ was determined by:

$$\boldsymbol{F}_g(\boldsymbol{x}) = (\rho(\boldsymbol{x})-\rho_v)\boldsymbol{g} \tag{15}$$

where $\boldsymbol{g}=(0,-g)$. Consequently, the total force in Eq. (13) is $\boldsymbol{F}=\boldsymbol{F}_m+\boldsymbol{F}_{ads}+\boldsymbol{F}_g$.

In this study, the energy equation was utilized to solve the temperature field, and it was discretized via finite difference method. Neglecting the viscous dissipation, the governing equation of the heat transfer can be determined as [18]:

$$\partial_t T = -u\cdot\nabla T + \frac{1}{\rho C_V}\nabla\cdot(\lambda\nabla T) - \frac{T}{\rho C_V}\nabla\cdot u\left(\frac{\partial P_{EOS}}{\partial T}\right)_\rho \tag{16}$$

where $\lambda$ is the thermal conductivity, and $C_V$ is the specific heat at constant volume. The right side of Eq. (16) is marked by $K(T)$. With the help of the fourth-order Runge-Kutta scheme [25], The time discretization of the temperature field was solved by the follow equation:

$$T^{t+\delta t} = T^t + \frac{\delta t}{6}(h_1+h_2+h_3+h_4) \tag{17}$$

where $h_1$, $h_2$, $h_3$ and $h_4$ can be determined as:

$$h_1=K(T^t), h_2=K(T^t+\frac{\delta t}{2}h_1), h_3=K(T^t+\frac{\delta t}{2}h_2), h_4=K(T^t+\delta t h_3) \tag{18}$$

In general, the flow process solved by MRT-LB model and the temperature field solved by FDM can be coupled by the $P_{EOS}$ in Eqs. (11) and (16). Moreover, the parameter $\rho$ and $u$ are determined by the pseudopotential MRT-LB model. The flow chart of the pseudopotential MRT-LBM model's simulation process can be found in Fig. 1.

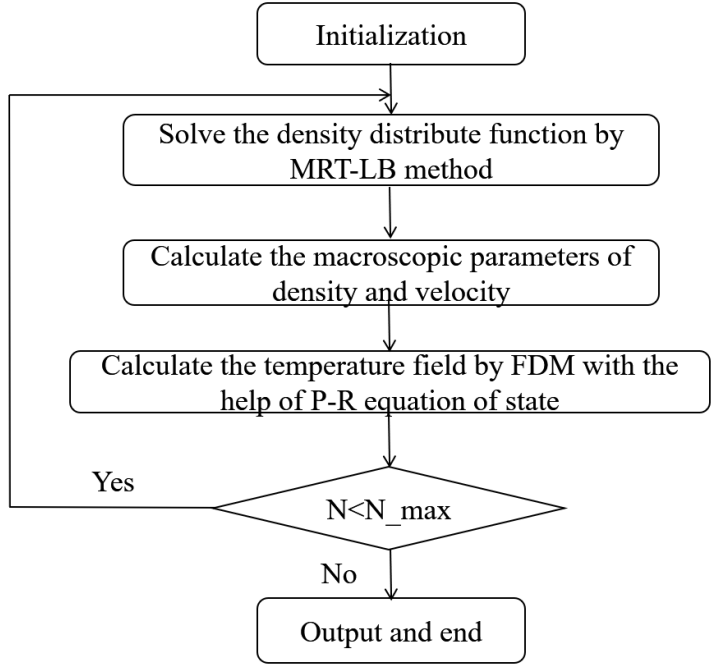

*Fig.1. Flow chart of the simulation process for the pseudopotential MRT-LB model*

## 3. Results and discussions

*3.1. Model verification*

To verify the pseudopotential MRT-LB model, the $D^2$ law for single water droplet evaporation was adopted [26]. In the computational domain, a single droplet was placed in the center of grid size $Nx \times Ny = 200 \times 200$ (lattice unit), the initial diameter of the droplet was assumed as D=60, and the temperature of the droplet and the saturated vapor near its surrounding was chosen as the saturation temperature $T_{sat}$=0.86$T_c$, which corresponded to the liquid density $\rho_l \approx 6.5$ and the vapor density $\rho_g \approx 0.38$. But the environmental temperature was assumed to be $T_c$, therefore the single water droplet placed in the center of the domain will start to evaporate due to the temperature gradient. In the verification simulation, all the material properties were assumed to be constants, except for the thermal conductivity, where two cases were used to demonstrate the effect from thermal conductivity to the evaporation: Case A: $\lambda=1/3$; Case B: $\lambda=2/3$.

As shown in Fig. 2, the diameter of the water droplet was decreasing as time marching due to evaporation. As the same time, at the same time instant, the larger the thermal conductivity is, the faster the evaporation happens. Fig. 3 provides the relation between the droplet diameter and time, where Dl denotes the diameter at different time instants, while D is the original diameter before evaporation. Obviously, there is a linear relation between $(Dl/D)^2$ and time, so this comparison demonstrates a perfect agreement with the $D^2$ law [26], it also provides us enough confidence to apply this model to oscillating multiphase heat transfer process in porous media.

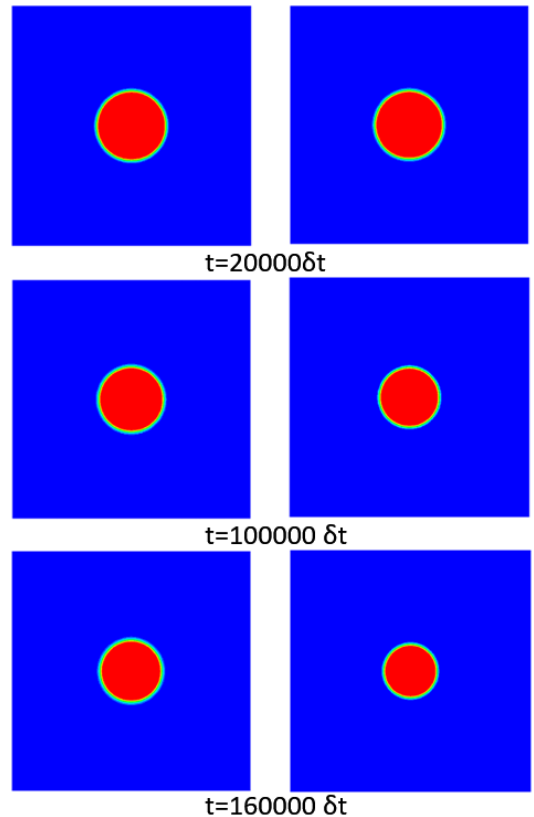

*Fig.2. Time evolution of single droplet evaporation with two different thermal conductivities.*
*Case A: λ=1/3(left); Case B: λ =2/3(right)*

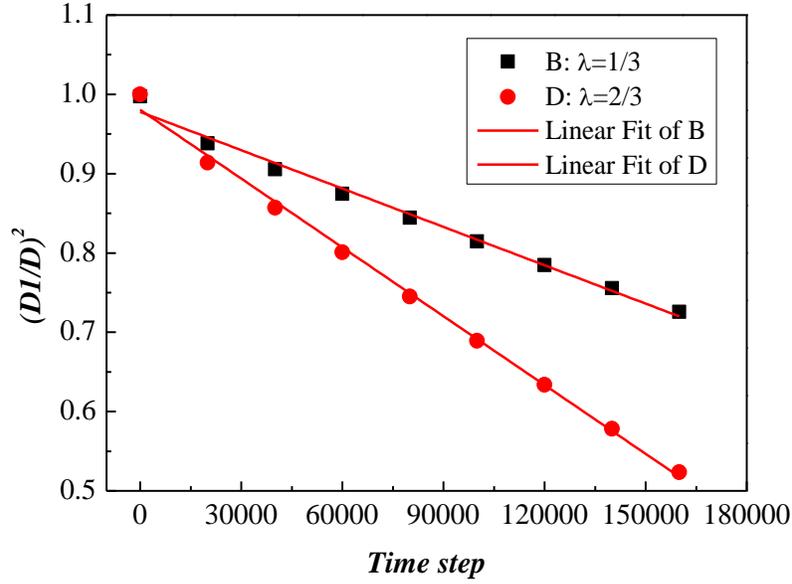

*Fig.3. Verification of the pseudopotential MRT-LB model using the D² law*

*3.2. Condensation of saturated water vapor along horizon wall under the gravity*

In this section, we will investigate the condensation of water vapor along horizontal walls under gravitational effects without oscillating flow. The computation domain is shown in Fig. 4, where the center of top boundary was assigned with lower temperature, while the two red top boundaries were assumed to be adiabatic. The bottom boundary was set as isothermal, and the left and right boundaries are periodic boundary conditions. The domain had a grid size of $Nx \times Ny = 100 \times 300$ (lattice unit). Furthermore, the top and bottom boundaries were all assumed as hydrophobic walls. Initially, the entire computational domain was filled with saturated water vapor, and the top boundary with lower temperature was set as $Tw = 0.65 Tc$, and both top and bottom boundaries were assumed as non-slip and adopted half step rebounding format. The specific heat capacity $C_v$ and thermal conductivity $\lambda$ of gas phase and liquid phase were assumed as 5 and 0.25, respectively. The gravity was set as $g = 0.0001$ (lattice unit).

It is important to note that the latent heat $h_{fg}$ can be determined through the model proposed in Ref. [27],

$$h_{fg} = \frac{a\alpha(T) + ac_\alpha \sqrt{T/Tc}\left[1 + c_a(1-\sqrt{T/Tc})\right]}{2\sqrt{2}b} \times$$
$$\ln\left|\frac{(b\rho_g - 1 - \sqrt{2})(b\rho_l - 1 + \sqrt{2})}{(b\rho_g - 1 + \sqrt{2})(b\rho_l - 1 - \sqrt{2})}\right| + ps(\upsilon_g - \upsilon_l) \quad (19)$$

and $h_{fg}$ can further be nondimensionalized by Jacob number,

$$Ja = \frac{C_v(T_{sat} - T_w)}{h_{fg}} \quad (20)$$

where $T_w$ is the wall temperature, $T_{sat}$ is the saturation temperature, and $C_v$ is the specific heat at constant volume.

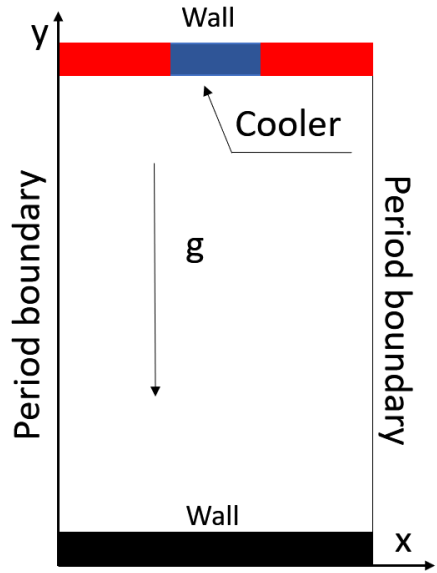

*Fig.4. Computational domain for the condensation of saturated water vapor on horizontal walls under gravity effect*

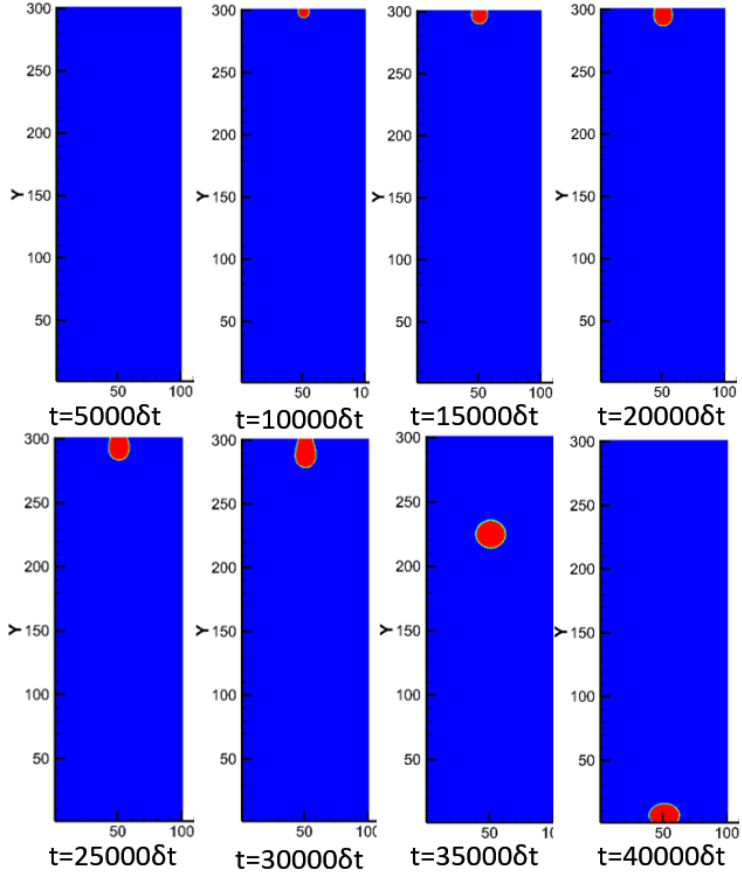

*Fig.5. Contour of gas/liquid phases during the entire lifecycle of saturated vapor condensation with vertical gravity*

Fig. 5 provides the entire life cycle of water vapor condensation on local cold wall and the free fall of droplet under gravity. Initially, there is no droplet forming on the top cold wall, as time marching the saturated water vapor starts to condensate on the local cold surface and forms a small droplet (t=10000 $\delta t$ ), then the droplet starts to develop and grow (t=15000 $\delta t$ , 20000 $\delta t$ , 25000 $\delta t$ ). Once the droplet becomes big enough, due to the combined effect from gravity and surface tension, the droplet is stretched along the negative Y-axis direction (t=30000 $\delta t$ ). Finally, the condensate droplet starts to fall under gravity (t=35000 $\delta t$ ) and impacts on the bottom boundary (t=40000 $\delta t$ ). Consequently, the pseudopotential MRT-LB model can accurately simulate the condensation process of saturated water vapor under gravity. In the next section, the oscillating flow and heat transfer will be discussed for condensation under gravity.

## 3.3. Condensation of oscillating saturated water vapor along horizontal walls with temperature gradient under gravity

The condensation of saturated water vapor with oscillating boundary condition was explored in this section. In this particular case, assume there is a stack of parallel plates, and the inlet boundary is experiencing oscillating pressure, while condensation of saturated vapor is expected to be discovered on the surface of parallel plates. To simplify the LBM simulation of condensation inside of 2D stack of parallel plates, one typical 2D channel with both top and bottom plates will be separated for investigation, and only the fluid domain will be studied. As shown in Fig. 6, the left boundary is the inlet of oscillating saturated water vapor, while the right boundary is assigned with constant pressure. The top and bottom walls are assumed with a temperature gradient, where the left side has higher temperature than the right side. At the same time, gravity is assumed to point to the right hand side. The top and bottom walls were assumed to be hydrophobic, the contacting angle was 137.82º, since the parameter $Gw$ in Eq. (14) was assumed to be 0.2. In the simulation, the grid size was assumed as $Nx \times Ny = 1600 \times 48$ (lattice unit), and the Zou-He boundary condition was adopted in this study. At the left boundary, the oscillating pressure was set as $p = p_0 + A\sin(2\pi ft)$, where $A = p_0/10$, and the inlet temperature of saturated water vapor was assumed as $T_{sat}$. The right boundary was set to be open to atmosphere, therefore the pressure was set as $p_0$. Both top and bottom walls were assumed as non-slip boundaries, and the temperature gradient was chosen as a linear distribution varying from T=$0.86T_c$ at left hand side to T=$0.56T_c$ at right hand side, where $T_c$=0.1096, calculated by Eq. (11). The material properties were the same as Section 3.2, for example the specific heat capacity $C_v$ and thermal conductivity $\lambda$ of gas phase and liquid phase were assumed as 5 and 0.25, respectively. The gravity was set as $g = 0.0001$ (lattice unit).

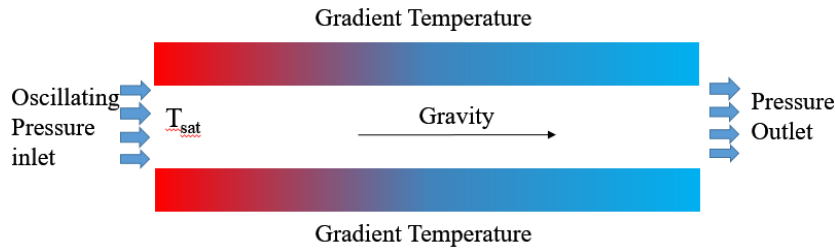

*Fig.6. Schematic of computational domain with oscillating inlet boundary and temperature gradient on the hydrophobic walls*

Fig. 7 illustrates the condensation process at various instants, for example at the exit of the computational domain, due to the existence of lower temperature ($<T_{sat}$), two tiny water droplets were formed on both top and bottom walls (t=8500$\delta t$), then one additional small droplet was generated in the upstream of the two existed droplets, which actually absorb more condensate and grow in a bigger size (t=9000$\delta t$). Under the effect of gravity, the droplets gradually move toward to the exit, and they further absorb the condensate on the two cold walls by forming larger size (t=9500$\delta t$, 10000$\delta t$). Finally, the two droplets, which are closest to the exit, merge together to form a bigger size one, and then eventually move out of the computational domain (X=1550) due to gravity, but two more tiny droplets are forming in the upstream again. Essentially, Fig. 6 tells us that the oscillating multiphase heat and mass transfer can keep feeding the exit region by forming larger size droplets on hydrophobic surface, as long as the condensate can be removed out from the channel, the oscillating flow and heat transfer will not be influenced.

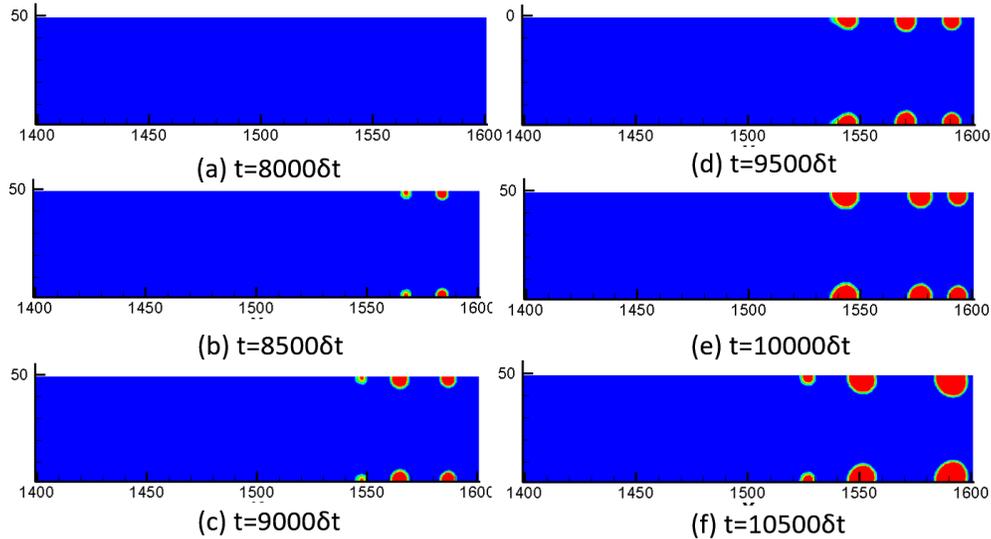

*Fig.7. Contour of two phases at different instants during the entire condensation process*

To further demonstrate the oscillating effect, Fig. 8 shows the temperature contour at the same time instants as shown in Fig. 7. It is important to point out that the condensation process is an exothermic process, therefore it is expected to find some regions in the temperature contour with higher temperature where the condensation is happening. Fortunately, Fig. 8 is showing excellent agreement with Fig. 7, because we can identify the exact locations where the condensation happens.

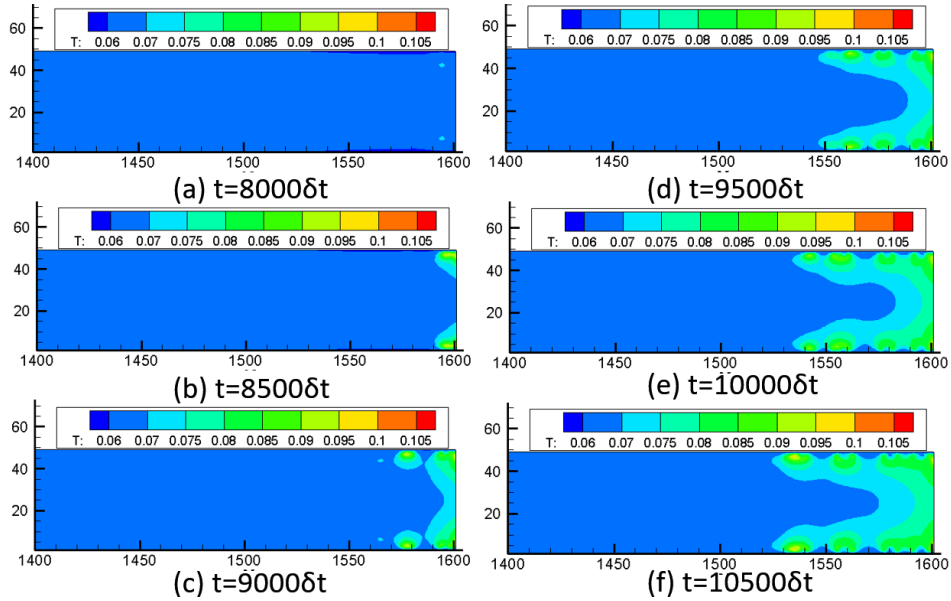

*Fig.8. Temperature contour of two phases during the life-cycle of saturated water vapor condensation*

To further demonstrate the effect of oscillating flow to the initial stage of condensation process in parallel plates, Fig. 9 shows the velocity vector at different time instants. At t=8000$\delta t$, due to the oscillating effect, the saturated vapor was driven to the right exit of the domain, even though there was no condensation appeared, the saturated water vapor started to accumulate at the right end, which was beneficial to initiate the condensation process. At t=8500$\delta t$, once the first droplet generated on the plate due to condensation, because the vapor flow was still in the first half cycle of the sinusoidal wave and moved to the right, more vapor was gathering around the condensed droplets, therefore the droplets started to grow and moved to the exit due to the gravitational effect, at the same time the velocity vectors clearly showed the external flow over those droplets. At t=9000$\delta t$, two smaller droplets were generated in the upstream of the two existed droplets, but the saturated vapor flow started to alter its direction to the left hand side of the domain because of its second half cycle of the sinusoidal wave, in other words there were more saturated vapor came into the domain from the right exit, consequently the two droplets close to the right exit became even bigger size, and the velocity vector also clearly showed the curvature of the flow path and its nearby flow field.

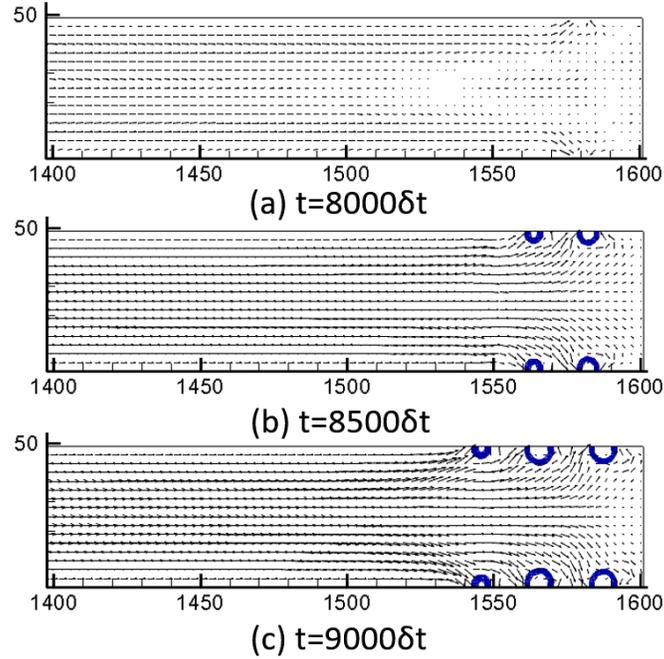

*Fig.9. Velocity vectors of saturated water vapor in the initial stage of condensation process at different time instants*

Because the wettability of the walls has a strong impact to the condensation process, two more cases were chosen to discuss the effect of wettability for oscillating flow: Case (a): CA=115º; Case (b) CA=145º, where CA is the contact angle. The different contact angles were realized by varying $G_w$ in Eq. (14), other parameters were maintained the same in Fig. 6.

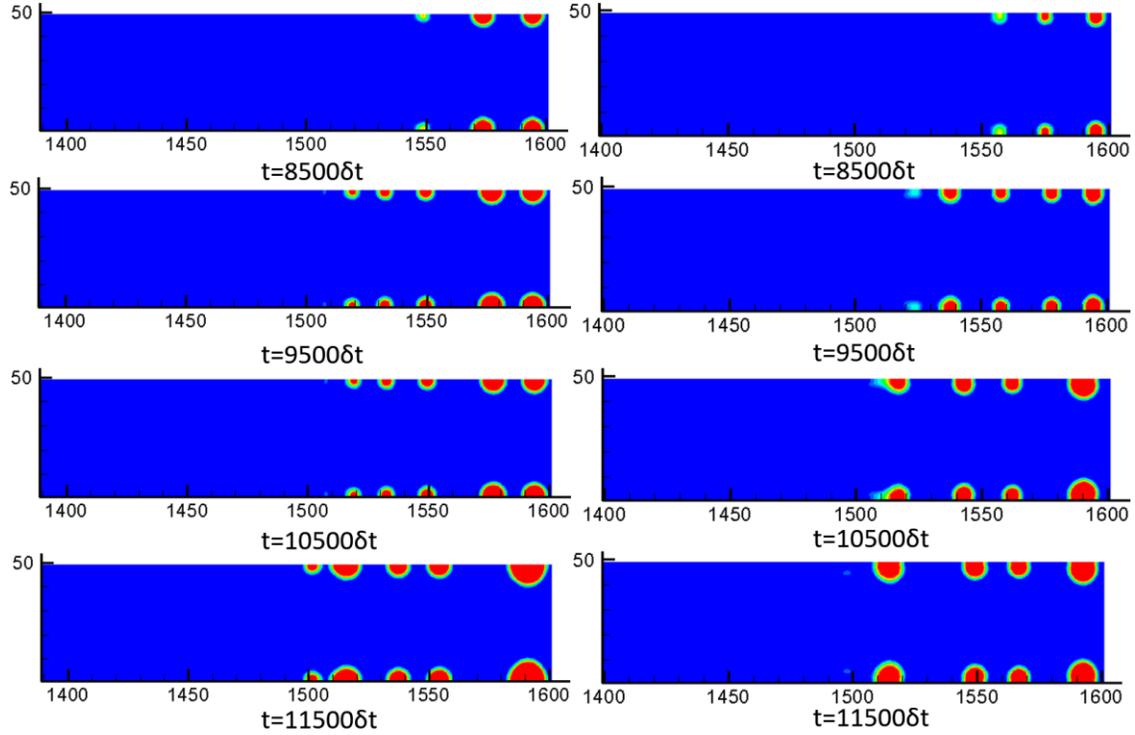

*Fig.10. Contour of two phases at different instants during the condensation with two different contact angles: Case (a): CA=115º (left column); Case (b): CA=145º (right column)*

Fig. 10 shows the contour of two phases at different time instants with different contact angles, where the left column is for CA=115º, and the right column is for CA=145º. Obviously, with the increasing of wettability, saturated water vapor was easier to condense on the cold walls, and the volume of condensed droplets was greater than the one with larger contact angle at t=8500$\delta t$. Furthermore, at t=10500$\delta t$ there were more condensed droplets formed on the surface with higher wettability (lower contact angle) within 1500<X<1600, but the distance between droplets was different for these two cases, because Case (a) had smaller spacing between each pair of droplets compared with Case (b), but the droplets for Case (b) were easier to move to the right end under the gravity, therefore it had more chances to condense more saturated water vapor and form more droplets.

## 4. Conclusions

Due to the importance of multiphase heat and mass transfer in porous media, especially the condensation in the thermal stack of thermoacoustic refrigerator, it is of great significance to understand the oscillating flow and condensation process inside the parallel plates. In the current study, a pseudopotential MRT-LBM model was adopted, and the model was verified by utilizing a single droplet evaporation case, and the results perfectly matched with the D$^2$ law, and then a continuous flow was adopted to further verify the model, by simulating the condensation on cold surface with vertical gravity. Finally, an oscillating flow and condensation in two parallel plates was simulated, where the saturated water vapor was condensing on the hydrophobic surface of plates. The following key findings can be concluded:

1) As long as the condensate can be removed from the parallel plates, the oscillating flow will not be influenced;
2) During the initial stage of condensation, the oscillating flow is actually beneficial to accumulate the saturated vapor at the exit regions, and the flow path around the droplets can also be observed;
3) Increasing the wettability will condense the water vapor more easily, and the large contact angle will form large spacing between each pair of droplets.

It's expected that this study can be used to optimize and redesign the structure of thermal stack in order to produce more condensed water, also this multiphase approach can be extended to more complicated 3D structures.

## NOMENCLATURE

| | |
|---|---|
| $a$ | constant in P-R equation of state |
| $b$ | constant in P-R equation of state |
| $c$ | lattice speed ( m/s ) |
| $c_s$ | lattice sound speed |
| $C_V$ | specific heat at constant volume ( J Kg$^{-1}$ K$^{-1}$ ) |
| $f$ | density distribution function |
| $f^{eq}$ | equilibrium density distribution function |
| $F$ | interaction force of particles ( N ) |
| $g$ | gravitational acceleration ( m/s$^2$ ) |
| $G_w$ | fluid-solid interaction strength |
| $h_{fg}$ | latent heat ($J/Kg$) |
| $Ja$ | Jacob number |
| $m$ | velocity moments |
| $m^{eq}$ | velocity equilibrium moments |
| $M$ | velocity orthogonal transform matrix |
| $Nx$ | lattice domain along along $x$ direction ($m$) |
| $Ny$ | lattice domain along along $y$ direction ($m$) |
| $p$ | pressure ($pa$) |
| $s$ | velocity diagonal relaxation time matrix |
| $\bar{S}$ | forcing term in the moment space |
| $t$ | time ($s$) |
| $T$ | temperature |
| $u$ | macroscopic velocity ($m/s$) |
| $u_x$ | velocity component along $x$ direction |
| $u_y$ | velocity component along $y$ direction |
| $\Delta x$ | length steep |

*Greek symbols*

| | |
|---|---|
| $\delta_t$ | time step |
| $\delta_x$ | lattice space length |
| $\lambda$ | thermal conductivity |
| $\theta$ | constant number |
| $\rho$ | density ($Kg/m^3$) |
| $\tau_\upsilon$ | kinematic viscosity of the fluid |
| $\Lambda$ | diagonal relaxation matrix |

*Subscripts*

| | |
|---|---|
| $ads$ | adhere wall force |
| *c* | critical state |
| *g* | gas phase |
| *i* | move direction of single-particle |
| *l* | liquid phase |
| *sat* | saturation state |
| *s* | solid |
| *w* | wall |
| *x* | Cartesian coordinate components |
| *y* | Cartesian coordinate components |

## ACKNOWLEDGMENTS

The authors are grateful for the support from New Faculty Supporting program (NFSP) at University of Texas Rio Grande Valley. The authors also appreciate the student financial support from National Natural Science Foundation of China (No. 11562011, No. 51566012), Graduate Innovation Special Foundation of Jiangxi Province (No. YC2017-S056) and Jiangxi Provincial Department of Science and Technology (No. 2009BGA01800).

## REFERENCES

[1] Hiller, R. A., & Swift, G. W. (2000). Condensation in a steady-flow thermoacoustic refrigerator. The Journal of the Acoustical Society of America, 108(4), 1521-1527.

[2] Slaton, W. V., Raspet, R., Hickey, C. J., & Hiller, R. A. (2002). Theory of inert gas-condensing vapor thermoacoustics: Transport equations. The Journal of the Acoustical Society of America, 112(4), 1423-1430.

[3] Rott, N. Thermoacoustics. Adv. Appl. Mech. 1980, 20, 135–175.

[4] Swift, G.W. Thermoacoustics: A Unifying Perspective for Some Engines and Refrigerators; Acoustical Society of America: Melville, NY, USA, 2002.

[5] Yu, G., Dai, W., & Luo, E. (2010). CFD simulation of a 300 Hz thermoacoustic standing wave engine. Cryogenics, 50(9), 615-622.

[6] Gunstensen, A. K., & Rothman, D. H. (1993). Lattice-Boltzmann studies of immiscible two-

phase flow through porous media. Journal of Geophysical Research: Solid Earth, 98(B4), 6431-6441.

[7] X. Shan, H. Chen, Simulation of nonideal gases and liquid-gas phase transitions by the lattice Boltzmann equation, Phys Rev E Stat Phys Plasmas Fluids Relat Interdiscip Topics, 49 (1994) 2941-2948.

[8] X. Shan, H. Chen, Lattice Boltzmann model for simulating flows with multiple phases and components, Phys Rev E Stat Phys Plasmas Fluids Relat Interdiscip Topics, 47 (1993) 1815-1819.

[9] Swift, M. R., Osborn, W. R., & Yeomans, J. M. (1995). Lattice Boltzmann simulation of nonideal fluids. Physical review letters, 75(5), 830.

[10] He, X., Chen, S., & Zhang, R. (1999). A lattice Boltzmann scheme for incompressible multiphase flow and its application in simulation of Rayleigh–Taylor instability. Journal of Computational Physics, 152(2), 642-663.

[11] He, X., Shan, X., & Doolen, G. D. (1998). Discrete Boltzmann equation model for nonideal gases. Physical Review E, 57(1), R13.

[12] Izham, M., Fukui, T., & Morinishi, K. (2011). Application of regularized lattice Boltzmann method for incompressible flow simulation at high Reynolds number and flow with curved boundary. Journal of Fluid Science and Technology, 6(6), 812-822.

[13] Zhou, W., Loney, D., Fedorov, A., Degertekin, F., & Rosen, D. (2013). Lattice Boltzmann simulations of multiple droplet interactions during impingement on the substrate. In The 24th Annual International Solid Free Form Fabrication Symposium Austin, TX, Aug (pp. 12-14).

[14] H.N.W. Lekkerkerker, C.K. Poon, P.N. Pusey, A. Stroobants, P.B. Warren, Lattice BGK Models for Navier-Stokes Equation, Europhysics Letters, DOI (1996) 479-484.

[15] P. Lallemand, L.S. Luo, Theory of the lattice boltzmann method: dispersion, dissipation, isotropy, galilean invariance, and stability, Physical Review E Statistical Physics Plasmas Fluids & Related Interdisciplinary Topics, 61 (2000) 6546-6562.

[16] M.E. Mccracken, J. Abraham, Multiple-relaxation-time lattice-Boltzmann model for multiphase flow, Physical Review E Statistical Nonlinear & Soft Matter Physics, 71 (2005) 036701.

[17] Q. Li, Y.L. He, G.H. Tang, W.Q. Tao, Improved axisymmetric lattice Boltzmann scheme, Physical Review E Statistical Nonlinear & Soft Matter Physics, 81 (2010) 056707.

[18] Q. Li, Q.J. Kang, M.M. Francois, Y.L. He, K.H. Luo, Lattice Boltzmann modeling of boiling heat transfer: The boiling curve and the effects of wettability, International Journal of Heat & Mass

Transfer, 85 (2015) 787-796.

[19] Q. Liu, Y.L. He, Q. Li, W.Q. Tao, A multiple-relaxation-time lattice Boltzmann model for convection heat transfer in porous media, International Journal of Heat & Mass Transfer, 73 (2014) 761-775.

[20] Q. Li, K.H. Luo, X.J. Li, Lattice Boltzmann modeling of multiphase flows at large density ratio with an improved pseudopotential model, Phys Rev E Stat Nonlin Soft Matter Phys, 87 (2013) 053301.

[21] Q. Li, K.H. Luo, Achieving tunable surface tension in the pseudopotential lattice Boltzmann modeling of multiphase flows, Physical Review E Statistical Nonlinear & Soft Matter Physics, 88 (2013) 053307.

[22] Q. Li, K.H. Luo, Thermodynamic consistency of the pseudopotential lattice Boltzmann model for simulating liquid–vapor flows, Applied Thermal Engineering, 72 (2014) 56-61.

[23] P. Yuan, L. Schaefer, Equations of state in a lattice Boltzmann model, Physics of Fluids, 18 (2006) 329.

[24] Q. Li, K.H. Luo, Q.J. Kang, Q. Chen, Contact angles in the pseudopotential lattice Boltzmann modeling of wetting, Phys.rev.e, 90 (2014) 053301.

[25] H. Liu, A.J. Valocchi, Y. Zhang, Q. Kang, Phase-field-based lattice Boltzmann finite-difference model for simulating thermocapillary flows, Phys Rev E Stat Nonlin Soft Matter Phys, 87 (2013) 013010.

[26] H. Safari, M.H. Rahimian, M. Krafczyk, Extended lattice Boltzmann method for numerical simulation of thermal phase change in two-phase fluid flow, Physical Review E Statistical Nonlinear & Soft Matter Physics, 88 (2013) 013304.

[27] S. Gong, P. Cheng, Lattice Boltzmann simulation of periodic bubble nucleation, growth and departure from a heated surface in pool boiling, International Journal of Heat & Mass Transfer, 64 (2013) 122-132.